# Magnetic Vortex with Skyrmionic Core in Thin Nanodisk of the Chiral Magnets

HAIFENG DU, WEI NING, MINGLIANG TIAN [(a)], AND YUHENG ZHANG

*High Magnetic Field Laboratory, Chinese Academy of Science, Hefei 230031, Anhui, P. R. China and Hefei National Laboratory for Physical Science at The Microscale, University of Science and Technology of China, Hefei 230026, People's Republic of Chinay*

Abstract –A type of vortex spin texture with skyrmionic core and a series of circle spin stripes was obtained in thin nanodisk of the chiral magnets with Dzyaloshinskii-Moriya interaction and uniaxial anisotropy by means of micromagnetic approach and Monte Carlo simulation. The size of skyrmionic core can be modulated continuously by controlling the disk size. Moreover, in some certain values of the disk size, this vortex state may be spontaneous ground state even without the help of the external magnetic field and thermal fluctuation. In addition, the uniaxial anisotropy is able to stabilize this vortex spin texture. We anticipate that the present work will inspire further experimental studies for exploring the spin structure.

Nanoscale magnetic elements are always a fascinating topic in last two decades motivated by its rich physics and relevance in a variety of miniaturized spintronics device [1]. With the prospect of yielding high symmetry and reproducible spin textures for data storage or other applications, particular attention has been given to high symmetry geometries. This has been demonstrated in nanodisks of the soft magnetic material, in which the magnetic vortex states, characterized by an in-plane curling magnetization around and an out-of-plane magnetization in the core, is very stable as the ground state in a wide range of disk size [2-4]. The formation of the vortex states in soft-magnetic nanodisk is due to the competing interactions between the ferromagnetic exchange and magnetostatic coupling. Within this standard model the vortex states are expected to be four-fold degenerate with two possible orientations of the rotation sense of the curling in-plane magnetization and polar direction of the perpendicular magnetization in the core [5,6].

However, recent experimental observations using full-field magnetic transmission soft X-ray microscopy confirmed that the vortex state in nanodisks experienced symmetry breaking mainly due to the anti-symmetry Dzyaloshinskii–Moriya (DM) interaction [7], which results from the broken inversion symmetry at surfaces or interfaces of thin magnetic films, as well as low-symmetry crystals. This chiral DM interaction was shown to favour non-uniform magnetic structures [8-11]. Prominently, the nowadays so-called skyrmion, a topologically stable vortex-like spin texture with unconventional spin–electronic phenomena, has been proved to exist in helical magnets with B20 structure [12]. But, in bulk materials, skyrmions only appeared in a tiny pocket in the temperature-magnetic field (T-H) plane. In contrast, stabilized skyrmions are able to survive over a relatively wide range in the T-H plane when the dimensionality of the helimagnet is reduced from its bulk to two dimensional (2D) thin films [13,14]. The occurrence of the highly stabilized skyrmions in thin film inspired the fabrication of 2D homogeneous thickness-controllable skymionic films by evaporation methods, which provides a good basis to engineer nanoscale pattern of helimagnets for miniaturized skyrmions-based spintronics device [15-17]. Unfortunately, the study of helimagnets in nanoscale is still in the infancy, and remains unexploited accordingly insofar. As a theoretical advance, the influence of the induced DM interaction on the vortex states in soft magnetic nanodisks has been investigated by Butenko *et al.,* where the DM coupling can increase or suppress sizes of vortices depending on its chirality [18].

In present paper, we theoretically investigated the magnetic microstructures in thin nanodisk of chiral magnets with DM interaction and uniaxial anisotropy by means of micromagnetic approach. A new type of vortex-like spin texture, described by a skyrmionic core with a series of circle spin stripes was obtained with free boundary conditions by minimizing the energy of the system [8,9], when the size of the nanodisk reaches a threshold value. The size of the skyrmionic core display quasi-periodic behaviour with the increase of the radius of the nanodisk. These properties were further confirmed by the Monte Carlo simulation. More importantly, these vortex states may be spontaneous ground state even without the help of the external magnetic field or thermal fluctuation at some values of nanodisk size. In

[(a)]E-mail: tianml@hmfl.ac.cn
[(a)]Present address: Hefei Institutes of Physical Science, CAS, 230031, Hefei, China

addition, uniaxial anisotropy is able to stabilize this vortex texture. A signature of such spiral spin texture, similar to the molecular arrangement with CF-2 type in liquid crystal [19], was preliminarily observed previously in 2D helimagnet FeGe [20]. Such a controllable spin topology might be useful in observing unconventional magneto transport properties.

A generally accepted simplest Hamiltonian for thin disk of the chiral magnets with DM interaction and uniaxial anisotropy is written as [21]

$$w = A(\nabla \boldsymbol{m})^2 - K_u m_z^2 - 1/2\, \boldsymbol{m} \cdot h_m + w_D \qquad (1)$$

where $\boldsymbol{m}$ is the unit vector of magnetization, $A$ is the ferromagnetic exchange constant; $K_u$ is the effective anisotropy constant with the easy axis perpendicular to the plane of the disk; $h_m$ is the stray field. In thin samples and for simplicity, the magnetostatic energy approximately reduces to a shape anisotropy $2\pi m_s^2$ with saturation magnetization $m_s$. Then an effective uniaxial anisotropy constant $K$ is defined as $K = K_u - 2\pi m_s^2$; and the last term $w_D$ represents the DM interaction, which is determined by the crystallographic symmetry. In this paper we use $w_D = D\boldsymbol{m} \bullet (\nabla \times \boldsymbol{m})$, arising from the cubic non-centrosymmetric helimagnets such as FeGe, $Fe_xCo_{1-x}Si$, and MnSi.

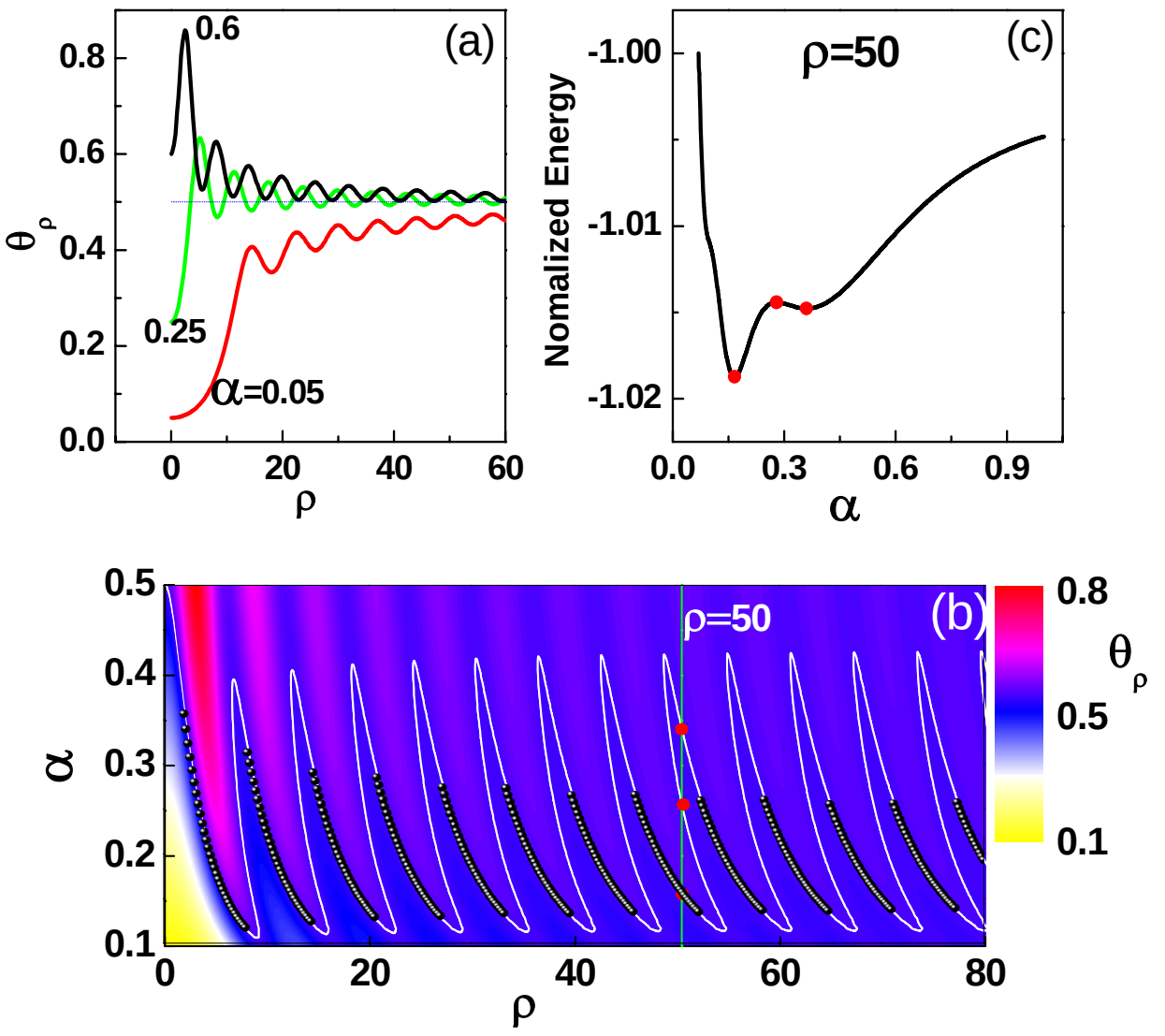

Fig.1.(a): Several typical curves of $\theta_\rho$ for different initial values $\alpha$; the boundary condition $\theta_\rho(R) = 0.5$ is shown as the dotted lines for comparison; (b): phase diagram for $\theta_\rho(\rho,\alpha)$, the dotted lines represent the required solutions $\theta_\rho(R) = 0.5$, three red intersection points between lines $\rho = 50$ and $\theta_\rho(\rho) = 0.5$, indicating multivalue behavior of required solutions; the final initial values $\alpha$, correspond to the global minimal energy of the systemdisplayedas black spheres; (c): The normalized energy as function of α, the three extreme points are relative to the three intersection points in (b).

Following the method well developed by Bogdanov [8,9,18] *et al.*, introducing spherical coordinates $\boldsymbol{m} = (\sin\theta\cos\psi, \sin\theta\sin\psi, \cos\theta)$ for magnetization and cylindrical coordinates for the spatial variable $\boldsymbol{r} = (\rho\cos\varphi, \rho\sin\varphi, z)$, minimizing functional (1) include axisymmetric localized solution $\psi = \psi(\varphi)$, $\theta = \theta(\rho)$. In cubic helimagnets, $\psi = \varphi + \pi/2$.

With the reduced length unit $A/D$, energy unit $D^2/A$, and uniaxial anisotropy constant $\beta = K/(D^2/A)$, the vortex energy density $E$ of the disk with radius $R$ can be expressed as $E = (2/R^2)\int_0^R w(\rho)\rho d\rho$, with

$$w(\rho) = \left[(d\theta/d\rho)^2 + \sin^2\theta/\rho^2\right] - \beta\cos^2\theta - (d\theta/d\rho + \sin\theta\cos\theta/\rho) \qquad (2)$$

Eq.(2) has been intensively investigated with several different boundary conditions. For 2D film of the chiral magnets, the boundary conditions are written as $\theta(0) = k\pi$ $\theta(\infty) = 0$, with integer number $k$ for isolated $k\pi$ vortex, and $\theta(0) = 0$, $\theta(\infty) = \pi$ for an isolated vortex [8,9]. In the case of small magnetic disk, the boundary condition is determined by the edge pinning effect imposed by the surface energy at the lateral disk edge. In soft magnetic nanodisks with induced DM interaction, the vortex core size is independence on the boundary conditions, thus, a free boundary condition is written as $\theta(0) = 0$, $\theta_\rho(R) = 0$ with $\theta_\rho = d\theta/d\rho$ [18]. In the thin disk limit, we neglected the edge pinning effect. However, our numerical results show that the free boundary conditions for soft nanodisks are not in consistence with the global minimal energy of the present system. The same arguments happen for other boundary conditions mentioned above.

The Euler equation for $\theta(\rho)$

$$d^2\theta/d\rho^2 + \theta_\rho/\rho - \sin\theta\cos\theta/\rho^2 - \sin^2\theta/\rho^2 - \beta\sin\theta\cos\theta = 0 \qquad (3)$$

In the thin disk, there are no constrained conditions on the edge. Even so, the extreme value of the energy density requires the natural boundary conditions, which are deduced from the equation $d(w(\rho)\rho)/d\theta_\rho|_{\rho=R} = 0$, thus

$$\theta(0) = 0 \quad , \quad \theta_\rho(R) = 0.5 \qquad (4)$$

Eq.(4) belongs to two point boundary value problem in the ordinary differential equations, which at the starting point do not determine a unique solution to start with. A random choice among the solutions that satisfy these incomplete starting boundary conditions is almost certain not to satisfy the boundary conditions at the end point $\theta_\rho(R) = 0.5$.

To solve the two point boundary value problem, akin to the method in ref.23, we firstly consider the standard initial value problem

$$\theta(0) = 0, \qquad \theta_\rho(0) = \alpha \qquad (5)$$

In this case, Eq.(3) may be solved just by numerical iteration from initial point to its end. For different values of $\alpha$,

the trajectories $\theta_\rho(\rho,\alpha)$ with $0<\alpha<\infty$ define a family of solutions, which include the required solutions with boundary conditions Eq.(4). Typical solutions $\theta_\rho(\rho,\alpha)$ for three different values $\alpha$ are shown in Fig.1.(a). These oscillation curves with period $T_\rho \approx 4\pi$ approach the value $\theta_\rho = 0.5$ with the increase of the radius of the disk. However, for $\alpha$=0.05 and 0.6, there is no intersection points between these curves and the line $\theta_\rho = 0.5$. When $\alpha$>0.6, the numerical results from random choosing value $\alpha$ support $\theta_\rho(\rho) > 0.5$, which implicates that the required solution exits only in certain interval of values $\alpha$. Moreover, we plot the phase diagram $\theta_\rho(\rho,\alpha)$ in the range of $0.1 \le \alpha \le 0.5$, as illustrated in Fig.1 (b), from which the required solutions with $\theta_\rho = 0.5$ are displayed as the white line. Obviously, the data demonstrated a multivalue behavior that several initial values $\alpha$ all satisfy the boundary condition Eq.(4) for a fixed disk size. In other words, Eq.(4) is only the necessary condition for global minimal energy of the system. For instance, Fig.1 (c) plots the normalized energy as the function of $\alpha$ for a fixed disk size $\rho = 50$. The initial values $\alpha$ (the red solid dots) which correspond to three extreme values of energy with two minima and one maximum are in consistence with three intersection points (the red solid dots) between the line $\rho = 50$ and white one $\theta_\rho = 0.5$ in Fig.1 (b). The required initial value $\alpha$, related with the global minimal energy of the system, is chosen by a Monte Carlo (MC) method. The yielding values $\alpha$ are shown as black spheres in Fig.1 (b). In this way, Eq.(3) with boundary conditions in Eq.(4) is solved.

To compare the present vortex state with the helix one, We also consider the energy density $E_h$ of the helix state [8,9], which is written as $E_h = (1/T_s)\int_0^{T_s} w_h(\rho)\,d\rho$, with

$$w_h(\rho) = \left(\frac{d\theta}{d\rho}\right)^2 - \beta\cos^2\theta - \frac{d\theta}{d\rho} \qquad (6)$$

where $T_s$ is the period of the helix structure, the Euler equation for $\theta(\rho)$ has the first integral

$$\left(\frac{d\theta}{d\rho}\right)^2 = \beta\sin^2\theta + c \qquad (7)$$

where the integration constant $c$ can be solved by an implicit equation

$$\int_0^\pi \sqrt{\beta\sin^2\theta + c}\,d\theta = \pi/2 \qquad (8)$$

Separating the variables and integrating (7) the $T_s$ is expressed as

$$T_s = \int_0^\pi 1\Big/\sqrt{\beta\sin^2\theta + c}\,d\theta \qquad (9)$$

In addition to the numerical method, Monte Carlo simulation is also used to confirm the numerical results. In the Monte Carlo simulation, a discrete 2D cubic lattice model based on the energy Eq.(1) is used to calculate the magnetic structure [22]. The disk is divided into a series of blocks with suitable lattice spacing $a$. Here, $a$ was chosen to yield the wavelengths $d = 20$ lattice constants. High temperature annealing Metropolis algorithm with free boundary condition is used to calculate the equilibrium spin microstructure of a thin disk. Since we concerned the vortex state, the magnetic moment in each block is not complete free, but fixed with direction perpendicular to the position vector $\boldsymbol{r}$ of the block due to constrain $\psi = \varphi + \pi/2$. Equilibrium situation is obtained by $2\times10^5$ Monte Carlo steps in each temperature. After a simple scalar transform, the simulated results are able to compare with the numerical ones.

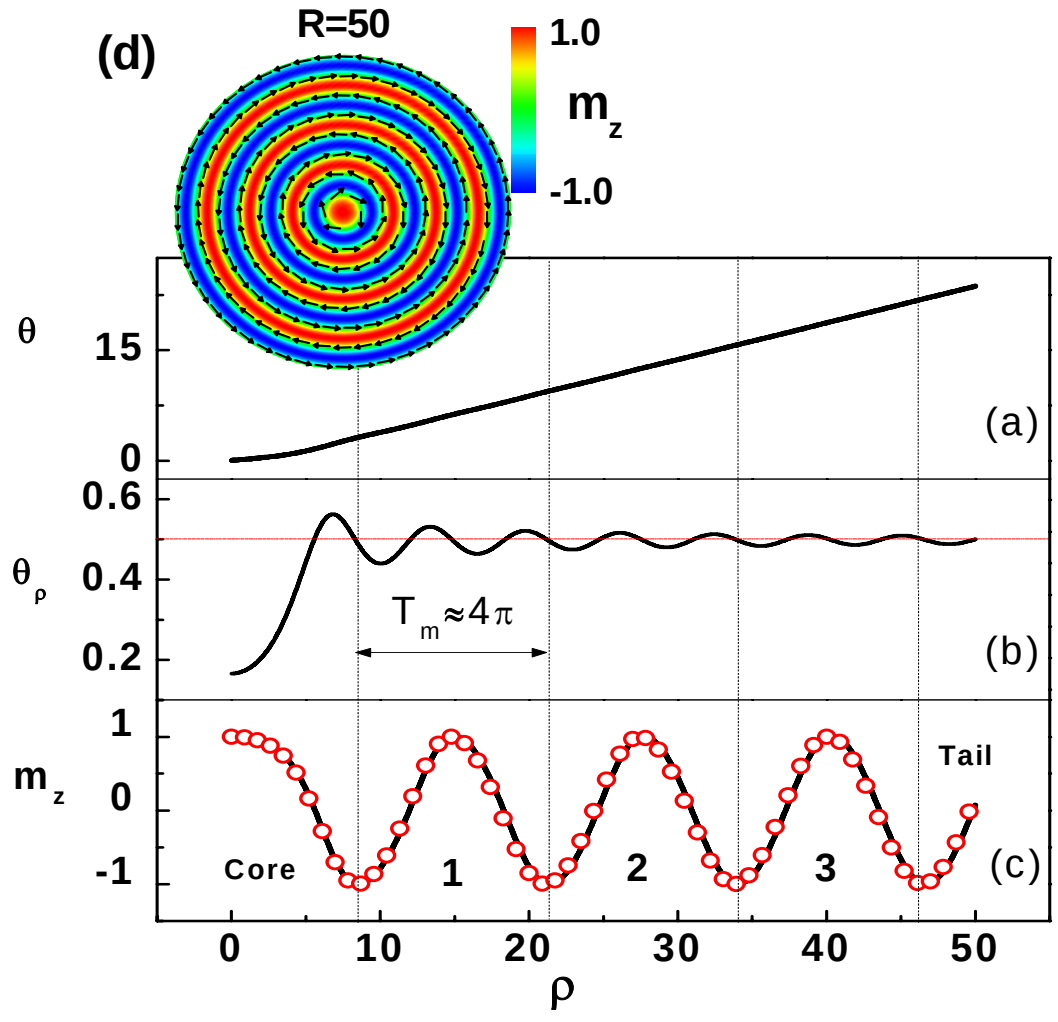

Fig.2.(a) and (b): Typical solution of $\theta$ and $\theta_\rho$ for a fixed radius $R = 50$; (c): the corresponding polar magnetization $m_z(\rho)$; the red empty circles is the results by Monte Carlo simulation; (d): the magnetization profile of the disk, where the spin texture is characterized by a skyrmionic core and a serial of circle spin stripes.

We firstly consider the case without the uniaxial anisotropy $\beta = 0$. Before discussing the numerical results, let's firstly give a few words on the asymptotic behavior of Eq.(3). When $\rho \to \infty$, Euler equation Eq.(3) for vortex structure approaches the one for helix case $d^2\theta/d\rho^2 = 0$ [9]. According to Eq.(4), Eq.(3) has the first integral $\theta_\rho = 0.5$, which is confirmed by the accurate numerical calculation, as shown in Fig.2 (a) and 2 (b), in which the curves $\theta(\rho)$ and $\theta_\rho(\rho)$ for disk radius $R = 50$ display approximate linear dependence with a slope of 0.5 and an oscillation behavior around $\theta_\rho = 0.5$ with a period of $T_{\theta_\rho} \approx 2\pi$, respectively.

The typical magnetization profiles for fixed radius $R = 50$ are illustrated in Fig.2 (c) and 2 (d). A significant feature presented here is that the spin-texture is characterized by a vortex core with a series of circle spin stripes. The period of stripes is nearly $T_m \approx 4\pi$, which approximately equals to that of the helix state. It must be noted that the final spin stripe (here, we called it tail with size $R_t$) is only partial owing to the limitation of the natural boundary conditions Eq.(4). The magnetic structure suggested above is well reproduced by Monte Carlo simulation, shown in Fig.2 (c) as red open circles.

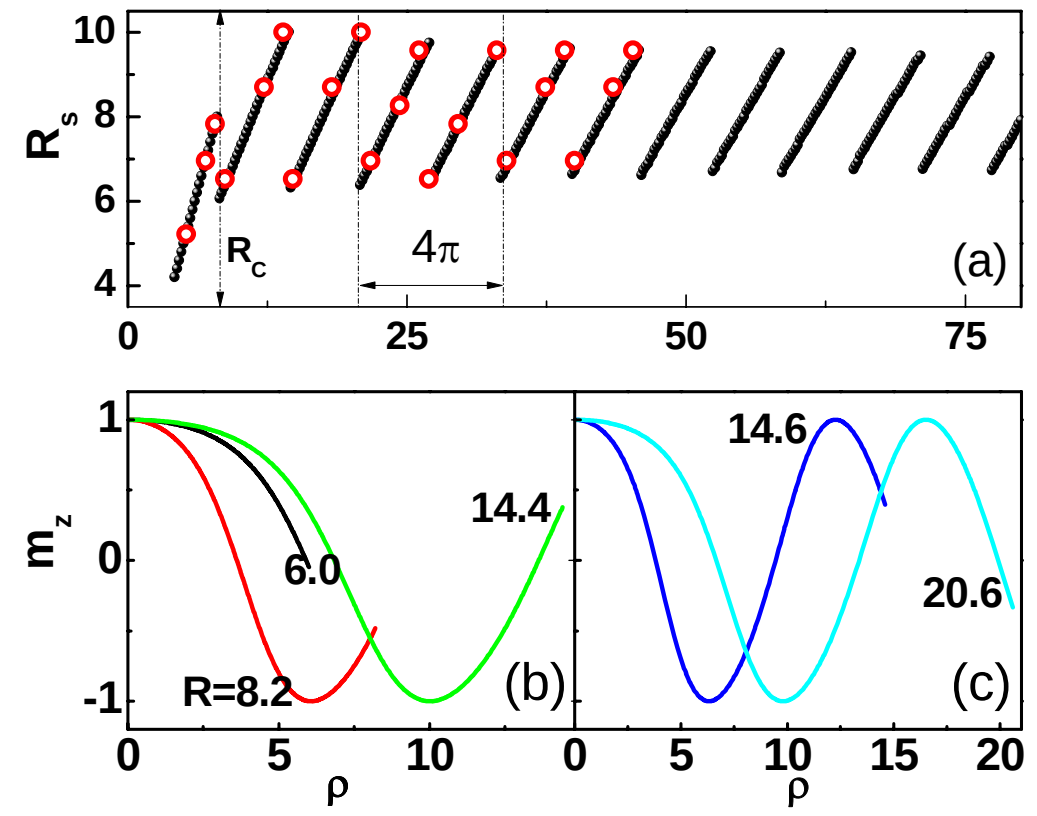

Fig.3. (a): The radius $R_S$ of skyrmionic core as a function of disk radius $R$, respectively; the red empty circles are the results by Monte Carlo simulation; (b) and (c): the polar magnetization profiles for different disk size.

The dependence of $R_S$ on the disk radius $R$ is displayed in Fig.3 (a). When $R < R_c$ with $R_c \approx 8.0$ as critical disk size, polar magnetization $m_z$ of thin disk shows monotonous change with the increase of radial distance $\rho$, as shown in Fig.3 (b) $(R = 6.0)$, in the edge of disk, $m_z \neq -1$. It is obvious that the vortex is not a skyrmion in this case. This result has an important implication for the future experiments by setting a limit on the skyrmion-based nanodevice. Just above the threshold value $R_c$, $R_S$ jumped to a small value, a complete skyrmion core with a tail size $R_t < 0.5T_m$ formed (Fig.3 (b), $R=8.2$). When the disk radius $R$ is lower than 14.4, both $R_S$ and $R_t$ increase linearly with increasing $R$ with the same slope 0.5 (Fig.3.(b), $R=14.4$). When $R>14.4$, $R_S$ jumped again to a small value and $R_t > 0.5T_m$ (Fig.3.(c), $R=14.6$), both $R_S$ and $R_t$ increase linearly with increasing $R_t$ up to 20.6 (Fig.3.(c)), above which an integral circle stripe appears with further increasing $R_t$. Curves $R_S(R)$ show the quasi-periodic behavior (also supported by MC simulation in Fig.3 (a) as red empty circle dots). These results suggest that, by controlling the size of the disk, one can realize the effective manipulation of the size of the skyrmionic core of thin chiral magnetic disk.

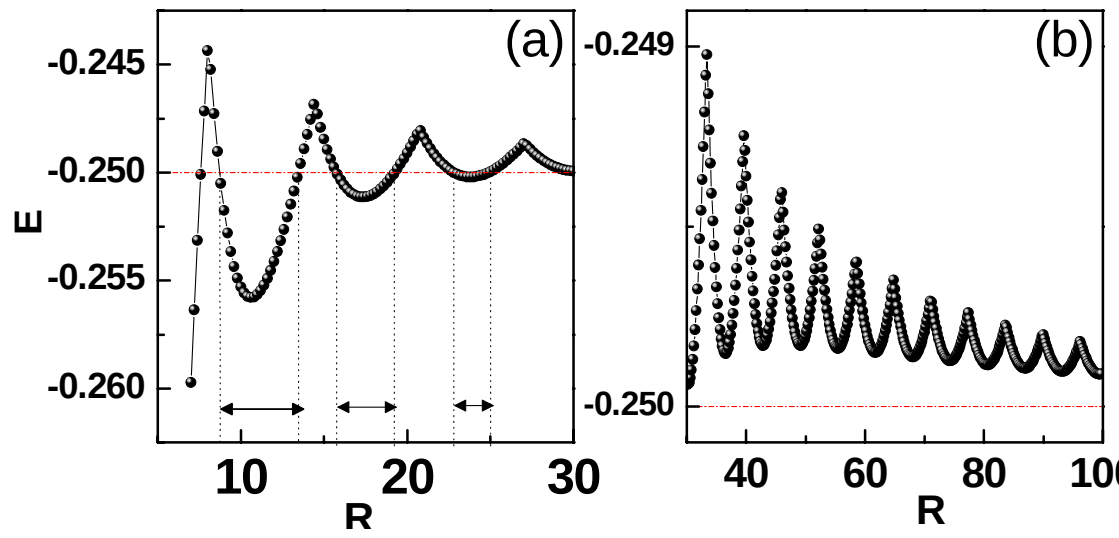

Fig.4 Energy density, E, as function of radius R of the disk, (a): $R \in (0,30)$; (b): $R \in (30,100)$. The red dotted lines represent the energy density of the helix state. In the intervals marked by the double arrows, the energy density of the vortex state is lower than that of the helix state, indicating a big possibility to form spontaneous vortex ground state even without the help of external field and thermal fluctuation.

The dependence of the energy density $E$ of the disk on its size $R$ is plotted in Fig.4. The red line $E_h = -0.25$ is the energy density of helix state. For clarity of the data, the horizontal axis is divided into two intervals, $R \in (0,30)$ and $R \in (30,100)$, respectively. In the former interval, $E(R)$ oscillated around $E_h$ with a quasi-period of $T_E \approx 2\pi$, indicating a possible vortex ground state since the energy of a vortex is lower than that of helix structure in some interval of disk radius $R$, marked by double arrows. Since the helix state would be distorted in the edge of disk due to the spatial confinement effect for magnetic disk, lowering the energy of the system, we couldn't allege that the spontaneous vortex ground state may form within this region, but it give a route to find this spontaneous vortex ground state even without the help of external field and thermal fluctuation with high probability. In the latter interval, the energy difference between the helix state and vortex state firstly increases, then, decreases with the increase of $R$. Especially, when $R>60.0$, the relative difference of energy between two states is only $0.1\%$, indicating a big chance to form this metastable vortex structure. In fact this circle spin stripe has been observed by Lorenz Microscopy in two dimensional chiral magnets FeGe in Ref.20 (Fig.1), where it is only partially displayed when one edge of thin sample is nearly round.

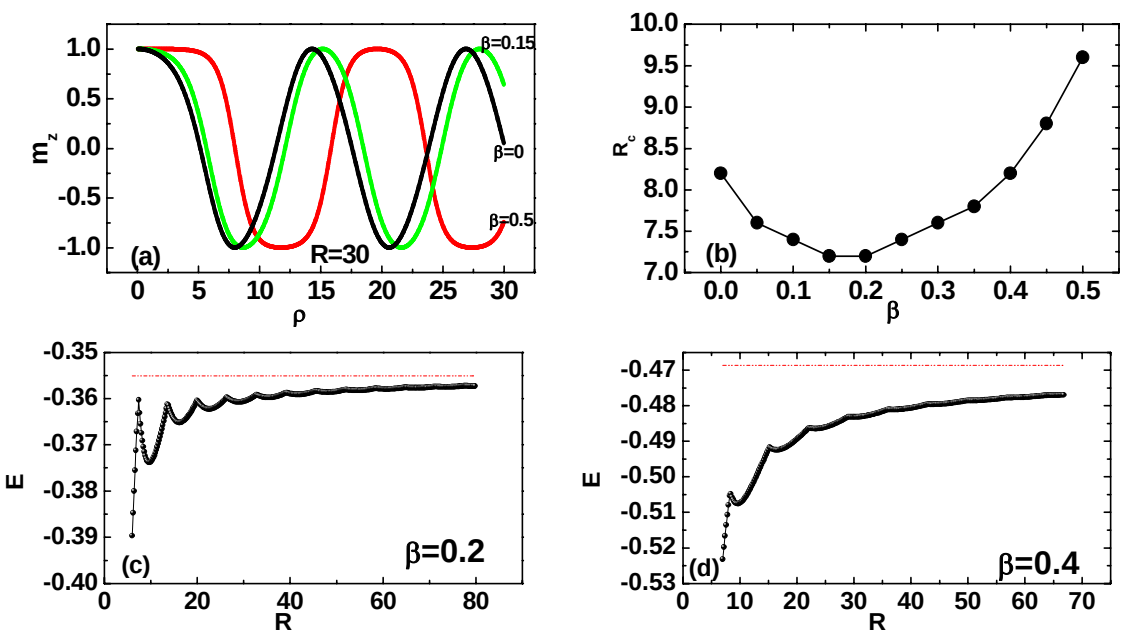

Fig.5. (a): the spin arrangement of the disk with different uniaxial anisotropy for radius $R=30$; (b): the critical radius, $R_c$, of the disk as the function of the uniaxial anisotropy $\beta$; (c) and (d) are, respectively, the energy density of the disk with reduced uniaxial anisotropy constants of $\beta \approx 0.2$ and $0.4$. The red dotted lines in the plots represent the energy density of the corresponding helix state.

In addition to two main energy scales, i.e. the ferromagnetic exchange and DM couplings, the uniaxial anisotropy energy, which arises from the strain or the symmetry-breaking in surface or interface of thin samples has been argued to response for the formation of skymionic ground states in thin layers $Fe_{0.5}Co_{0.5}Si$ [11]. Here, the uniaxial anisotropy has significant effect on the magnetic properties of vortex states. The spin arrangement of the disk with different uniaxial anisotropy for radius $R=30$ is shown in Fig.5 (a). With increasing uniaxial anisotropy, the region from $m_z = 1$ to $m_z = 1$ gradually becomes narrow, and the spin arrangement finally transforms into the stripe-like magnetic structure with chiral magnetic order. Similar magnetic structures has recently been observed in Fe 2ML/Ni/Cu(001) magnetic films with DM interaction and high uniaxial anisotropy [23]. In Fig.5 (b), we show the critical radius $R_c$ of the disk as a function of the uniaxial anisotropy $\beta$. When $\beta$ increases from zero, $R_c$ decreases and reaches a minimal value at $\beta \approx 0.2$. After that, $R_c$ increases with increasing $\beta$. The energy density $E$ of the disk with reduced uniaxial anisotropy constants $\beta \approx 0.2$ and $0.4$ is show in Fig.5 (c) and 5(d), respectively. With the increase of the uniaxial anisotropy, the energy difference increased between vortex state and helix one, indicating that the uniaxial anisotropy is able to stabilize this vortex state.

It is worthy comparing the current result with the previous one in Ref.18, where the influence of the DM interaction on the vortex states in soft magnetic nanodisks has been investigated. There, the easy plane magnetic material with the anisotropy constant $\beta < 0$ is considered in order to guarantee the soft magnetic vortex ground state. Meanwhile, the characteristic constant $D/\sqrt{AK}$ is generally chosen below the threshold value $4/\pi$ since above the value the magnetization of a layer would transform into a modulated spin textures, which are involved in the present paper.

In real helimagnets, some weak energy contributions, such as the dipolar energy and the crystal field energy, also play important roles in determining the properties of helimagnets. In analogous to the soft magnetic vortex in nanodots and the bubble domains in film, dipolar interaction may stabilize present vortex states since it is a magnetic charge free spin microstructure [3,9]. However, in one dimensional system, dipolar interaction leads to the magnetic easy axis of system along wire direction, which favours the helix state, as shown in the MnSi nanowire where only a signature of helimagnetic ordering is observed by magnetoresistance measurements [24]. Crystal field energy favours the spin along its easy axis direction, and thus promoting the helix state. This is why the vortex states are only found in FeGe with weak crystal field and the dislocation-like defects are often observed in $Fe_{0.5}Co_{0.5}Si$ with stronger crystal field energy [20]. Additionally, we have not investigated the magnetic dynamics of the nanodisk in spite of its importance, since the magnetization curves of helimagnets are generally calculated by circular cell approximation in 2D infinite film or bulk materials [8-10]. It is hard to adopt this numerical method to the finite-size objects.

We have suggested a type of spin vortex texture with skyrmionic core in a thin nanodisk of helimagnets by numerical calculations. The size of skyrmionic core can be tuned by controlling the disk size. Moreover, this special spin texture may form spontaneous ground state even without external field and thermal fluctuation in some certain interval of the disk size. In addition, the effect of the uniaxial anisotropy on the properties of disks including the spin arrangements, the critical disk radius and the energy of the vortex state is also discussed. Considering the importance of skyrmions at nanoscale materials, we anticipate that our present work will inspire further theoretical and experimental studies for exploring the magnetic properties of Skymions at nanoscale, and even fabricate miniaturized skyrmion-based device.

**ACKNOWLEDGEMENTS**: This work was supported by the National Key Basic Research of China, under Grant No. 2011CBA00111 and No. 2010CB923403; the National Nature Science Foundation of China, Grant No. 11174292 and No. 11104281，No. 11104280; and the Hundred Talents Program of the Chinese Academy of Sciences.